\documentclass[amsmath,amssym,amsfonts,pre,twocolumn,floatfix]{revtex4}
\usepackage[dvips]{epsfig}

\newcommand{\xv}{{\mathbf x}}
\newcommand{\rv}{{\bf r}}
\newcommand{\uv}{{\bf u}}

\newcommand{\kv}{{\bf k}}

\newcommand{\cH}{{\cal H}}

\newcommand{\grad}{{\bf \nabla}}

\begin{document}

\title{Structural Transitions and Soft Modes in Frustrated DNA Crystals}

\author{Gregory M. Grason}
\affiliation{ Department of Polymer Science and Engineering, University of Massachusetts - Amherst, MA 01003, USA}

\begin{abstract}
Relying on symmetry considerations appropriate for helical biopolymers such as DNA and filamentous actin, we argue that crystalline packings of mutually repulsive helical macromolecules fall principally into two categories:  unfrustrated (hexagonal) and frustrated (rhombohedral).   For both cases, we construct the Landau-Ginzburg free energy for the 2D columnar-hexagonal to 3D crystalline phase transition, including the coupling between molecular displacements {\it along} biopolymer backbone to displacements in the plane of hexagonal order.  We focus on the distinct elastic properties that emerge upon crystallization of helical arrays due to this coupling.  Specifically, we demonstrate that frustrated states universally exhibit a highly anisotropic in-plane elastic response, characterized by an especially soft compliance to simple-shear deformations and a comparatively large resistance to those deformations that carry the array from the low- to high-density crystalline states of DNA.
\end{abstract}

\date{\today}

\maketitle

Even ignoring its fundamental role as the carrier of genetic information for many living organisms, DNA is a remarkable molecule, which continues to draw significant study purely as a component of molecular systems with complex order.  Although negatively charged in aqueous solution, DNA can be condensed by a variety of methods~\cite{bloomfield, gelbart_bruinsma} yielding a remarkably rich spectrum of liquid-crystalline order.  Roughly speaking, as the concentration of condensed DNA is increases a sequence of thermodynamic phases can be observed:  isotropic $\rightarrow$ blue (or pre-cholesteric) phases $\rightarrow$ cholesteric $\rightarrow$ columnar hexagonal $\rightarrow$ crystalline~\cite{livolant_doucet, livolant_durand, livolant_leforestier}.  Of course, it is this latter crystalline state that is the focus of studies  which resolve the double-helical structure of DNA on the atomic scale~\cite{timsit_moras, berman}.  

The precise nature of the crystalline ordering of DNA is quite a complex issue stemming from the microscopic frustration introduced by the hexagonal packing of helical molecules.  While single helices can be uniformly packed, the perfect packing of symmetric bi-helical molecules, such as F-actin, cannot be achieved in a hexagonal lattice.  Molecular DNA is intermediate to these limiting cases although experimental observations of crystalline states suggest that interactions between DNA are highly frustrated.  It is the aim of this Letter to demonstrate that as a direct result of this molecular-level packing frustration, the 3D crystalline state exhibits an unusual response to shear in the plane of hexagonal order, characterized by a generic softness to simple-shear deformations.  

At the high packing densities of crystalline DNA, the microscopic interactions between neighboring helices are complex, involving a combination of electrostatic, hydrogen bonding, steric and ion-mediated forces.  Rather than relying on a detailed description of interactions at the molecular scale, we appeal to symmetry-based considerations to explore the consequences of frustration of helical molecules.  To this end, we construct an order-parameter expansion of the free energy near to the 2D columnar-hexagonal to 3D crystalline phase transition.  Based on generic symmetry considerations of  helical polymer arrays, we deduce that nascent crystalline groundstates fall largely into two categories:  unfrustrated or weakly frustrated possessing hexagonal symmetry; and highly frustrated with rhombohedral symmetry.  The minimal coupling allowed by symmetry between in-plane (bending modes) and out-of-plane (torsion or sliding modes) deformation dictates the unusual elastic properties of DNA groundstates.  In particular, for the case of highly frustrated groundstates, we find that the onset of order along the out-of-plane backbone direction dramatically alters the in-plane response of the array, leading to the emergence of a set of soft elastic modes in the crystal.  Remarkably, the elastic response is highly resistant to those lattice distortions that transform crystalline packing from one frustrated groundstate to another.  

\section{Interactions and groundstates of helices} 

\begin{figure}
\center
\epsfig{file=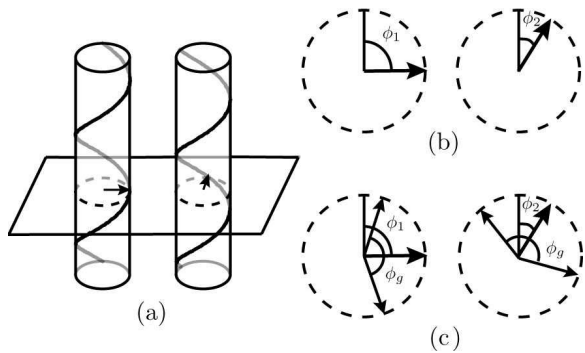, width=3in}
\caption{Following ref. \cite{manna} the orientation of helical molecules is defined by projecting the helical groove location onto a bi-secting plane: (a) side view and (b) top view.  The geometry of orientation for double-helical molecules is shown in (c), where $\phi_g$ is the angular width of the minor groove. }
\label{fig.1}
\end{figure}

We begin by considering the interactions between aligned, single-helical molecules and demonstrate that crystalline groundstates of helical molecules on closed-packed lattices belong to two classes, {\it unfrustrated} and {\it frustrated}.  In their most general form, the interactions between a pair of single helices, $i$ and $j$, can be written as,
\begin{equation}
\label{eq.1}
U_{{\rm sh}} = \sum_{n=0}^{\infty} U^{(n)}_{{\rm sh}} \cos \big[ n(\phi_i-\phi_j) \big] \ ,
\end{equation}
where $U^{(n)}_{{\rm sh}}$ are coefficients and $\phi_i$ and $\phi_j$ are defined.  Consider a horizontal plane intersecting the $i$th helix at some height, $z=0$ say [see Figure \ref{fig.1} (a)].  The azimuthal angle between vector connecting the central axis of the molecule to the spiral at that height and some reference direction (here $\hat{y}$) defines $\phi_i$.  Note that due to helical symmetry of the molecule, $\phi_i$ can be changed by simple rotation {\it or} translation of the molecule.  This special symmetry has been shown to the possibility of a peculiar, ``screw liquid crystalline" phase of helical molecules~\cite{manna}.

In general, $U^{(n)}_{{\rm sh}}$ appearing in eq. (\ref{eq.1}) depends on details of the inter-molecular forces present, helix separation, and solution conditions such as concentrations of depletants or various ions species.  Due to the finite-size molecule subunits of any helical molecules -- such as, bases or G-actin monomers for the respective cases of DNA and F-actin -- we can assume that series $|U^{(n)}_{{\rm sh}}|$ is monotonically decreasing.  This was shown explicitly for the case of the electrostatic interactions between uniformly charged helices~\cite{kornyshev_leikin}.  As such, we can assume that the $\phi$-dependence of interactions are dominated by first few terms of eq. (\ref{eq.1}).   Below, we will focus on the terms $n=1$ and $n=2$, as these seem sufficient for describing observed DNA packings.

Here, we avoid making specific assumptions about the nature of microscopic interactions.  Progress can nevertheless be made by considering two simple cases:  repulsive and ``sticky" (or attractive) helices.  Considering the geometry of Fig \ref{fig.1} it is clear that when $\phi_i-\phi_j = (2 m+1) \pi$ ($m$ is any integer) the strands of neighboring helices are in close contact.  For the case of repulsive (sticky) helices we therefore expect this close-contact orientation to correspond to a local maximum (minimum) of $U_{{\rm sh}}$.  On these grounds we should expect that for the case of repulsive interactions, $ U^{(n)}_{{\rm sh}}  < 0 $ when $n$ is odd and $ U^{(n)}_{{\rm sh}}  > 0 $ for even $n$.  Again, this expectation is borne out by electrostatic calculations of like-charged helices.  Of course, the opposite -- $ U^{(n)}_{{\rm sh}}  > 0 $ for $n$  odd and $ U^{(n)}_{{\rm sh}}  < 0 $ for even $n$ -- would be true for sticky helices.

For double-helical molecules like DNA, we can write the interaction in terms of inter-helix registry by summing over the interactions between single strands.  This yields an expression similar to eq. (\ref{eq.1}), with single-helix coefficients replaced with double-helix coefficients, $U^{(n)}_{{\rm dh}} = 2 U^{(n)}_{{\rm sh}} \cos^2(\phi_g/2)$, where $\phi_g$ is the azimuthal separation between helical strands [see Figure \ref{fig.1} (c)].  As required by symmetry, we see that for perfectly symmetric double-helical molecules, such as F-actin, with $\phi_g = \pi$, the odd $n$ terms of $U^{(n)}_{{\rm dh}}$ vanish.  B-form DNA with a minor groove width of roughly of $\phi_g \simeq 0.7\pi-0.8 \pi$, retains the $n=1$ term, although with considerably diminished amplitude.  This, along with considerations of the foregoing paragraphs, suggests that we may capture the essential features of DNA packing with the simplified interaction potential,
\begin{equation}
\label{eq.2}
U_{{\rm DNA}} \simeq - U_1 \cos [ \phi_i - \phi_j ]+U_2 \cos \big[2( \phi_i - \phi_j) \big] \ ,
\end{equation}
where $U_1$ and $U_2$ are positive coefficients and we neglect the constant $n=0$ term.  

\begin{figure}
\center
\epsfig{file=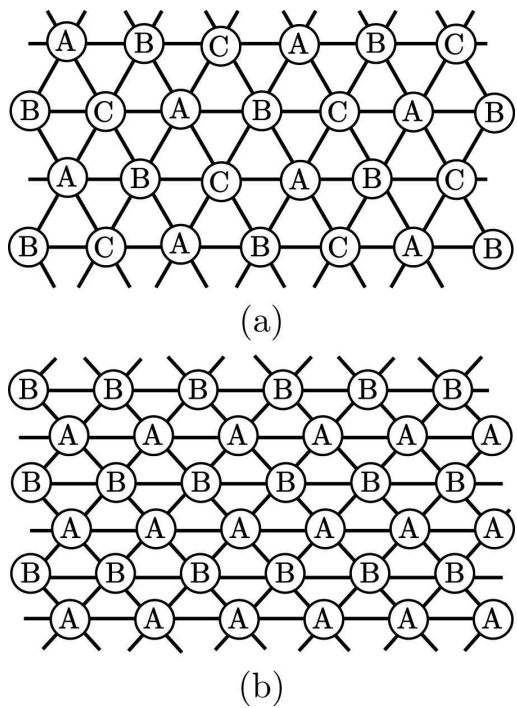, width=2.65in}
\caption{(a) shows the symmetry of the rhombohedral groundstate of helical packing.  The sublattices A, B and C refer relative rotational displacements of helices.  In the case of extreme frustration, sublattices B and C are respectively displaced $\pm  \pi /3$ (or $\mp \pi/3$) relative to A, where $p$ is the helical pitch.  (b) shows are alternative orthorhombic groundstate which alters the underlying hexagonal lattice of lower density columnar phase.  The relative displacement of sublattices A and B are dictated by details of the interhelical potential.}
\label{fig.2}
\end{figure}

Focusing, in part, on the detailed dependence of $U_1$ and $U_2$ electrostatic forces present for DNA in aqueous solution, Kornyshev and coworkers have demonstrated that interhelical potentials such as described by eq (\ref{eq.2}) lead to complex range of crystalline groundstates~\cite{harreis, cherstvy, wynveen, rmp}.  For the case that $U_1 \gg U_2$, interactions are effectively {\it ferromagnetic}, XY-type and unfrustrated, leading to ground states with uniform azimuthal orientation.  In the opposite limit, when $U_2 \gg U_1$, interactions are frustrated on close-packed, hexagonal lattices.  Defining the angle $\Phi_i = 2 \phi_i$, the second term in eq. (\ref{eq.2}) is essentially an {\it anti-ferromagnetic} XY-type coupling in $\Phi_i$, which cannot be mutually optimized be three pairs of adjacent molecules.  Although this term prefers $\Phi_i-\Phi_j=\pi$ for all neighboring pairs, the best that can be accomplished on the hexagonal lattice with are configurations with $\Phi_i$ changing by increments of $\pm 2 \pi/3$ around the triangular plaquettes~\cite{lee}.    Hence, the frustrated groundstate breaks the full six-fold rotational symmetry of the hexagonal lattice, retaining only a three-fold rotation symmetry [see Figure \ref{fig.2} (a)].  Careful experimental studies DNA condensates~\cite{livolant_durand, livolant_leforestier} reveal that close to the transition, crystalline structures maintain the hexagonal ordering of the lower density columnar-hexagonal phase.  Only at higher densities does the hexagonal order give way to an orthorhombic groundstate [Fig. \ref{fig.2} (b)].  In order to resolve this structural transition in detail it is necessary to invoke a specific microscopic description of separation-dependence of interaction parameters, $U_1$ and $U_2$, as in the electrostatic theory of refs. ~\cite{harreis, cherstvy, wynveen, rmp}.

An analysis of the groundstates of eq. (\ref{eq.2}) for helical molecules on the hexagonal lattice~\cite{cherstvy} reveals that the unfrustrated groundstate persists for $U_1>4U_2$.  Outside of this range, the frustrated groundstates consist of three staggered sublattices--labeled as $\xv_{\rm A}$, $\xv_{\rm B}$ and $\xv_{\rm C}$ respectively--as shown in Figure \ref{fig.2}.  If helices on sublattice A have orientation, $\phi_{\pm}(\xv_{\rm A})=\phi_0$, then those on the respective B and C sublattices are given by $\phi_{\pm}(\xv_{\rm B})= \phi_0 \pm \delta \phi$ and $\phi_{\pm}(\xv_{\rm C})=\phi_0 \mp \delta \phi$, where the value of $\delta \phi$ varies continuously from the unfrustrated state, $\delta \phi=0$, when $U_1=4 U_2$ to the fully-frustrated state, $\delta \phi = \pi/3$ when $U_2 \gg U_1$.  Here, $\pm$ refer to the two degenerate groundstates, which differ by a rotation in the plane by $2\pi/6$.  

Clearly, groundstate properties helical molecular assemblies are most sensitive to the precise values of $U_1$ and $U_2$ for DNA and other biopolymers, and these coefficients themselves will have a complex dependence on microscopic parameters resulting from a combination of intermolecular forces.  Even without a full knowledge of the inter-molecule forces at play in biopolymer systems, we can make progress by noting that groundstates of helical molecules will largely fall into two classes, frustrated and unfrustrated~\footnote{We note that observations of frustrated DNA crystals in which $\delta \phi \simeq 2 \pi/6$~\cite{livolant_leforestier} suggest the degree of frustration of helical interactions is considerable.}.  The symmetry of the former is hexagonal, while the latter possess only a rhombohedral symmetry.  In the remainder of the article we focus on the emergent elastic properties of these groundstates based on the broken symmetry of crystalline groundstates.  A similar symmetry-based approach was used by Lorman, Podgornik and \v{Z}ek\v{s} to explore the thermodynamics of the transitions between low-density hexatic phases and high-density orthorhombic phases of DNA~\cite{lorman}.  Unlike that scenario, we find below that it is not, in general, possible accomplish that transition {\it continuously} if the system first passes through an intermediate rhombohedral packing.

\section{Crystallization of unfrustrated groundstates}
Having considered the groundstate structure of helical arrays, we now turn to construct a proper description of transition from the 2D columnar-hexagonal phase to a higher density, 3D frustrated or unfrustrated crystalline phase.  We begin by constructing an order-parameter theory for the transition to the simpler unfrustrated groundstate, where helices have a uniform orientation (i.e. $\phi_i = {\rm const.}$ for all $i$).  

In the absence of the out-of-plane order, the columnar hexagonal phase is described by the displacement field $\uv (\rv)$ that describes fluctuations of the polymer backbone in the xy plane of hexagonal order.  Here, $\rv =\xv + {\bf z}$, where $\xv$ labels the in-plane position of the chain in the hexagonal lattice and ${\bf z}$ describes the position along the polymer backbone.  The (coarse-grained) elastic energy describing fluctuations of this two-dimensionally ordered phase can be written as ~\cite{selinger},
\begin{equation}
\label{eq.3}
\cH_\perp = \frac{1}{2} \int d^3 x \Big[ 2 \mu u_{ij}^2+ \lambda u_{kk}^2 + K_3 (\partial^2_z \uv)^2 \Big] \ .
\end{equation}
Here, $i$, $j$ and $k$ refer only to in-plane directions and $u_{ij}=(\partial_i u_j + \partial_j u_i)/2$ is the symmetric, 2D elastic strain.  The first and second terms describe distortions of the hexagonal packing and the third term describes the energy cost for fluctuations that bend the polymers.  In principle, this energy may also include additional penalties for so-called twist and splay deformations ~\cite{degennes}.  

In the unfrustrated groundstate, the onset of order along the polymer backbone direction can be described by the development of a non-zero amplitude of a smectic-like order parameter, $\psi(\rv)$~\cite{degennes}.  Specifically, we can define the order parameter, which represents modulations of the material density, $\rho(\rv)$ along the $\hat{z}$ direction, through $\rho(\rv) \simeq {\rm Re} \big[1+ \psi(\rv)e^{iG z}\big] \rho_{{\rm Hex}}(\rv) $, where $\rho_{{\rm Hex}}(\rv)$ is the density field of the 2D array of polymers. The special symmetry of helical molecules with pitch, $p$, is captured by the following transformation properties of the order parameter.  Under translation along $\hat{z}$ by $\Delta z$, $\psi \rightarrow e^{iG \Delta z} \psi$, where $G=2 \pi /p$.  And under rotation around $\hat{z}$ by $\Delta \theta$, $\psi \rightarrow e^{i\Delta \theta} \psi$.  This ensures 1) that the order parameter has the appropriate periodicity and 2) that a translation followed by the appropriate rotation leaves a helical 
groundstate unchanged.  

Since the order parameter of the unfrustrated groundstate is essentially that of a 3D XY model, the following Landau free energy describes the onset of a non-zero amplitude of $\psi(\rv)$ at the columnar hexagonal to crystalline transition, 
\begin{equation}
\cH_{|\psi|}=\frac{1}{2} \int d^3 \big[ r|\psi|^2+ \frac{u}{2} |\psi|^4 \big] \ .
\end{equation}
In addition to $\cH_{\|psi|}$ we must also consider gradient terms that are consistent with the symmetry of both phases.  On it's own eq. (\ref{eq.3}) has the symmetry properties of a 2D hexagonal array of chains, namely, the energy is invariant under rotations around $\hat{z}$ by $2 \pi /6$.  The unfrustrated groundstate maintains this in-plane symmetry, only breaking translational symmetry along the backbone direction.  To second order in derivatives we have the following contributions to the free energy of the array,
\begin{eqnarray}
\label{eq.5}
&&\!\!\!\!\!\!\!\!\!\!\cH_{\grad} = \frac{1}{2} \int d^3 x\Big\{ c_{\parallel} |\partial_z \psi|^2  + c_\perp\big|(\grad_\perp + i G \partial_z \uv)\psi|^2   \\
  &&\!\!\!\!\!\!\!\!+b u_{kk} {\rm Re}\big[ i \psi \partial_z \psi^* \big]+ |\psi|^2\big (2 \delta \rho~ u_{kk}+ \delta \lambda~u^2_{kk}+2 \delta \mu ~ u_{ij}^2 \big ) \Big\} . \nonumber
\end{eqnarray}
Here, $c_{\parallel}$, $c_{\perp}$, $b$, $\delta \rho$, $\delta \mu$ and $\delta \lambda$ are coefficients.   Note the appearance of the gauge coupling between in-plane gradients of $\psi$ and $z$ derivatives of backbone displacement in the second term of (\ref{eq.5}).  This is identical to the rotationally-invariant coupling between gradients of the order parameter and the director for smectic phases~\cite{degennes}, as $\partial_z \uv$ is the in-plane component of the backbone tangent vector.  It is this coupling which is responsible for the resistance of the crystal to shear deformations that slide molecules past another~\cite{grason}.

The specific critical properties of the phase transition from 2D columnar tho 3D solid order are unusually complex, owing to the gauge coupling between in-plane displacement modes and fluctuations of the smectic-like order parameter.  For the purposes of the present study we focus on emergent properties outside of the critical region.  In the crystalline state (above a critical density) $\psi$ takes on a non-zero amplitude, $|\psi_0|=(-r/u')^{1/2}$, for $r<0$, where $u'=u-\delta \rho/4(\lambda+\mu)$.  To examine fluctuations of the groundstate, we consider {\it phase} variations of the order parameter, $\psi(\rv)=|\psi_0| e^{i \phi(\rv)}$, which represent torsional or translational displacements of the helical macromolecules.  Defining $C_\parallel = c_\parallel|\psi_0|^2$, $C_\perp = c_\perp|\psi_0|^2$, $2B=b|\psi_0|^2$, $\lambda'=\lambda+|\psi_0|^2\delta \lambda$ and $\mu'=\mu+|\psi_0|^2\delta \mu$ we have the following effective Hamiltonian for in-plane ($\uv$) and out-of-plane ($\phi$) fluctuations,
\begin{eqnarray}
\label{eq.6}
 \cH_{eff} [\uv,\phi] \!\!&\!\!\!\!=\!\!\!\!&\!\! \frac{1}{2} \int d^3 x\Big[ C_\parallel(\partial_z \phi)^2+C_\perp(\grad_\perp \phi+G \partial_z \uv)^2 \\ \nonumber &&\!\!\!\! +2B u_{kk}~\partial_z \phi+2\mu'~u_{ij}^2+\lambda' u_{kk}^2 + K_3(\partial_z^2 \uv)^2 \Big]  .
\end{eqnarray}
Here, we will focus on the renormalized elastic behavior of the in-plane displacement modes due to the coupling between the out-of-plane modes appearing in the crystalline phase.  Integrating out the smooth (Gaussian) fluctuations of the $\phi$ field in (\ref{eq.6}), it is straightforward to derive an effective Hamiltonian (in Fourier space) for the $\uv$ fluctuations in the crystalline phase,
\begin{equation}
\label{eq.7}
 \cH'_{eff}[\uv] = \frac{1}{2} \int~\frac{d^3 k}{(2 \pi)^3}\big[S_L^{-1}(\kv) |\uv_L(\kv)|^2+S_T^{-1}(\kv) |\uv_T(\kv)|^2\big]  ,
\end{equation}
where $\uv_L=\kv_\perp (\kv_\perp \cdot \uv)/k_\perp^2$, $\uv_T=\uv - \uv_L$,
\begin{eqnarray}
\label{eq.8}
S_L^{-1}(\kv)  \!\!&\!\!\!\!=\!\!\!\!&\!\! (\lambda'+2\mu')k_\perp^2+K_3 k_z^4 \nonumber
\\ && +\frac{C_\parallel C_\perp G^2 k_z^4 - B(2C_\perp G+B)k_z^2 k_\perp^2}{C_\parallel k_z^2+C_\perp k_\perp^2} \ ,
\end{eqnarray}
and 
\begin{equation}
\label{eq.9}
S_T^{-1}(\kv) = \mu' k_\perp^2 +K_3 k_z^4+C_\perp G^2 k_z^2 \ .
\end{equation}
The transverse modes, $u_T(\kv)$, are generically stable and behave as simple phonons in an anisotropic 3D crystal, while the fluctuation spectrum of the longitudinal modes is considerably more anisotropic owing to the coupling to out-of-plane modes.

\section{Crystallization of frustrated groundstates}
Having analyzed the simpler unfrustrated case, we now construct the Landau-Gizburg approach to the columnar to crystalline transition when the groundstate is highly frustrated.  Here, the groundstates have the staggered structure shown in Figure \ref{fig.2} (a), with helical molecules occupying 3 sublattices.  The symmetry of the groundstates is encoded by the phases $\phi_{\pm}(\xv)$ discussed above, so that the order parameter for the frustrated states can now be written as,
\begin{equation}
\label{eq.10}
\psi(\rv)=\psi_+(\rv) e^{i \phi_+(\xv)}+\psi_-(\rv) e^{i \phi_-(\xv)}
\end{equation}
where again $\pm$ corresponds to crystalline ordering in either of the two degenerate possibilites.  Hence, for highly frustrated helical groundstates we have a 2-component complex vector order parameter $(\psi_+,\psi_-)$.

As both groundstates have rhombohedral symmetry, they are invariant under rotations in the plane by $2 \pi/3$, but they are not six-fold symmetric.  Consider a rotation in the plane of $2 \pi/6$ around an axis passing through the A sublattice.  From Fig. (\ref{fig.2}) it is clear this operation takes A sites to A sites and interchanges B and C sites.  Since $\phi_{\pm}(\xv_{\rm B}) = -\phi_{\pm}(\xv_{\rm C})$, under roations by $2 \pi/6$ in the plane, $(\psi_+, \psi_-) \rightarrow (\psi_-, \psi_+)$ (up to an overall phase shift due to helicity of the molecules).    In constructing the Landau-Ginzburg free energy for the transition, all terms must respect the six-fold rotational symmetry of the higher-symmetry, columnar hexagonal phase.  To fourth order in $|\psi|^4$ the most general Landau free energy describing this type of $U(1) \times Z_2$ symmetry breaking is~\cite{kawamura},
\begin{eqnarray}
\cH_{|\psi_{\pm}|} \!\!&\!\!\!\!=\!\!\!\!&\!\! \frac{1}{2} \int d^3\big[ r(|\psi_+|^2+|\psi_-|^2)  \nonumber \\ && +\frac{u}{2}(|\psi_+|^2+|\psi_-|^2)^2 +  v|\psi_+|^2|\psi_-|^2\big] .
\end{eqnarray}
Note that $\cH_{|\psi_{\pm}|}$ is invariant under interchange of $+$ and $-$ components (rotations by $2 \pi / 6$).  Spontaneous symmetry breaking is described by the case $r<0$, $u>0$ and $v>-2u$ ($v=-2u$,$ r=0$ is a tricritical point).  In this case, the order parameter assumes a non-zero amplitude in either the $+$ or $-$ component, while the other component goes to zero.  Here, this broken discrete symmetry indicates the broken six-fold rotation symmetry of the hexagonal columnar phase.

For the case of the frustrated groundstate, the same gradient terms appearing in eq. (\ref{eq.5}) for the {\it unfrustrated} case will also appear in $\cH_{\grad}$ for the {\it frustrated} case, except that each term with $\psi$ in the former case is replaced by term with both  $\psi_+$ and $\psi_-$, maintaining $+$ and $-$ exchange symmetry.  In addition to the gradient terms consistent with hexagonal symmetry, we are obliged to consider coupling terms that are consistent with the lower three-fold symmetry frustrated groundstates.  This term is constructed from products of the vectors,
\begin{equation}
v^{\pm}_j \equiv {\rm Re} \big[ i \psi_\pm (\grad_\perp+i G \partial_z \uv)_j \psi^*_\pm\big] ,
\end{equation}
and $u_{ij}$ which transform as a vector and tensor, respectively, in the xy plane.  Following ref. \cite{landau}, it can be shown that the three-fold rotational symmetry of the rhombahedral state allows for two possible free-energy contributions,
\begin{equation}
\lambda_x^{\pm}\big[(u_{xx}-u_{yy})v^{\pm}_x-2u_{xy}v^{\pm}_y\big]+\lambda_y^{\pm}\big[(u_{xx}-u_{yy})v^{\pm}_y+2u_{xy}v^{\pm}_x\big],
\end{equation}
where $\lambda_x^{\pm}$ and $\lambda_y^{\pm}$ are coefficients.  One of these two terms must be eliminated due to the symmetry properties of frustrated groundstates of helical molecules, which are invarient under the combined operation $y\rightarrow -y$ and $z\rightarrow - z$ (see the in-plane mirror symmetry in Fig. \ref{fig.2}).  Note, that the second reflection is required by the chiral symmetry of the molecules.  Since $v_i^{\pm}=-v_i^{\pm}$ under $z\rightarrow -z$, this requires that $\lambda_x^{\pm}=0$ as this term changes sign under the mirror reflections, leaving only $\lambda_y^{\pm} \neq 0$.  Of course, for a crystal of {\it achiral} molecules, the opposite scenario would be true ($\lambda_x^{\pm} \neq 0$ and $\lambda_y^{\pm}=0$).

To be clear, we have constructed elastic terms, $\lambda^{\pm}\big[(u_{xx}-u_{yy})v^{\pm}_y+2u_{xy}v^{\pm}_x\big]$ that respect the three-fold symmetry of the frustrated crystalline states, $\psi_\pm$.  Under rotations by $2\pi/6$, this term changes sign.  Since the $+$ and $-$ states are exchanged under this operation, choosing $\lambda^{+}=-\lambda^- \equiv \lambda_f$ yields a six-fold rotationally invariant contribution to the free energy,
\begin{equation}
\delta \cH_{\grad}= \lambda_f\int d^3 x \Big[(v_y^+-v_y^-)(u_{xx}-u_{yy})  + 2 (v^+_x-v^-_x)u_{xy}\Big] .
\end{equation}
As before we examine the effects of fluctuations of the out-of-plane and in-plane degrees of freedom in the broken symmetry crystalline state, by considering phase variations of the order parameter.  If we take as our broken symmetry state, say, $\psi_+(\rv)=|\psi_0| e^{i \phi (\rv)}$ and $\psi_-(\rv) =0$, we then have the following additional elastic contribution to the fluctuation Hamiltonian, $\cH_{eff}[\phi,\uv]$, for frustrated groundstates,
\begin{eqnarray}
\label{eq.15}
\delta \cH_{eff}[\phi, \uv] \!\!&\!\!\!\!=\!\!\!\!&\!\! \Lambda_f \int d^3 x\Big[(\partial_y \phi+G \partial_z u_y)(u_{xx}-u_{yy})\nonumber \\ && \ \ \ \ +2(\partial_x \phi+G \partial_z u_x)u_{xy}\Big] .
\end{eqnarray}
where $\Lambda_f=\lambda_f |\psi_0|^2$.  

\begin{figure}
\epsfig{file=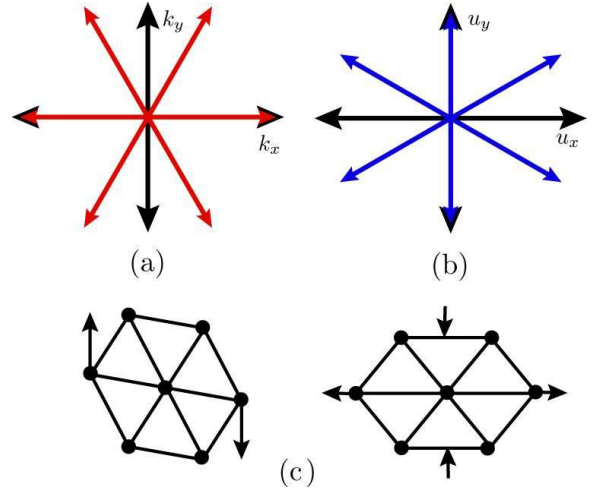,width= 3in}
\caption{The wavevectors and polarizations  of soft elastic modes of the frustration rhombohedral state are shown in (a) and (b) respectively.  In (c), the geometry of the soft simple shear modes is shown on the left, while the elongation shear that carries the rhombohedral state  [Fig. \ref{fig.2} (a)] to the higher density orthorhombic state [Fig. \ref{fig.2} (b)] .  }
\label{fig.3}
\end{figure}

The appearance of this new coupling between $\phi$ and $\uv$ modes in eq. (\ref{eq.15}) reflects a highly anisotropic elastic response of the frustrated crystalline state.  Integrating out the out-of-plane ($\phi$) modes yields expressions for the in-plane response of the system even more complex than eqs.  (\ref{eq.7})-(\ref{eq.9}).  We can nevertheless get a sense of the elastic response of the frustrated states by focusing on a simple limits of the full elastic response as defined by,
\begin{equation}
\cH'_{eff} =\frac{1}{2} \int \frac{d^3k}{(2 \pi)^3} \sum_{ij} u_i(\kv) S^{-1}_{ij} (\kv) u_j(-\kv) ,
\end{equation}
where $i$ and $j$ refer to independent directions in the plane.  The components of $S_{ij}^{-1}(\kv)$ can be extracted from eqs. (\ref{eq.6}) and (\ref{eq.15}).  Specifically, we focus on the $k_z=0$ terms in the limit of incompressibility, $\kv_\perp \cdot \uv(\kv)=0$.  For this special case, $S_{ij}^{-1}(\kv)$ is diagonal in the basis, $u_1(\kv)$ and $u_2(\kv)$ which can be written in terms of x and y displacements as,
\begin{eqnarray}
\label{eq.19}
u_1(\kv)=\frac{2 k_x k_y}{k_\perp^2}u_x(\kv) +\frac{(k_x^2-k_y^2)}{k_\perp^2}u_y(\kv)  , \\
u_2(\kv)=\frac{(k_x^2-k_y^2)}{k_\perp^2}u_x(\kv) -\frac{2 k_x k_y}{k_\perp^2} u_y(\kv) ,
\end{eqnarray}
with eigenvalues,
\begin{eqnarray}
\label{eq.21}
S_{11}^{-1}(\kv_\perp,k_z=0)=\Big(\mu'-\frac{\Lambda_f^2}{C_\perp}\Big)k_\perp^2 , \\  S_{22}^{-1}(\kv_\perp,k_z=0)=\mu' k_\perp^2 .
\end{eqnarray}
Here, we find an especially soft compliance along the $\uv_1(\kv)$ directions.  The shift of $S_{11}^{-1}(\kv_\perp,k_z=0)$ due to the coupling to out-of-plane $\phi$ modes is proportional to $|\psi_0|^2$, indicating the splitting between the response to the $\uv_1$ and $\uv_2$ modes grows with increasing crystallinity.  The incompressibility constraint, $\kv_\perp \cdot \uv=0$ implies a relation between the principle directions of in-plane deformation.  This constraint only allows for deformations with $\uv_1\neq0$ while $\uv_2 = 0$ when $k_y=0$ or when $3k_x^2=k_y^2$.  Hence, the wavevectors of these soft modes lies along the crystal axes of the underlying hexagonal lattice (see Figure \ref{fig.3}).  From eq. (\ref{eq.19}) we find that the polarization of the soft modes is perpendicular to its direction in $\kv$-space.  That is, upon crystallization to frustrated groundstate, the helical array is most easily deformed by {\it simple shear} with the shear direction perpendicular the hexagonal axes.

It must be emphasized that special compliance to simple shearing of the underlying hexagonal lattice of the frustrated state is a necessary consequence of the discrete breaking of 6-fold symmetry and does not rely on details of the microscopic interactions between DNA molecules.  This analysis relies on only 1) the 3-fold rotational symmetry of the frustrated crystalline groundstate 2) the underlying 6-fold symmetry of columnar phases and 3) the lack of mirror symmetry of the constituent chiral molecules~\footnote{We note, specifically, that the in-plane response would be qualitatively different for frustrated array of {\it achiral} filaments.}.  A direct signature of this anisotropic phonon dispersion appears in the diffuse thermal scattering surrounding Bragg peaks in the $k_z=0$ plane.  Near to a Bragg spot at reciprocal vector, ${\bf G}$, this scattering is encoded in $\langle |{\bf G} \cdot {\uv (\Delta \kv) }|^2 \rangle$, where $\Delta \kv$ is the wavevector deviation from ${\bf G}$.  This analysis predicts pronounced thermal scattering (for $\Delta \kv \perp {\bf G}$) around Bragg peaks whose ${\bf G}$ lies along one of the six soft-mode polarizations (Fig. \ref{fig.3}), and reduced thermal scattering for ${\bf G}$ off those axes.  %For ${\bf G}$ lying between these peaks, the diffuse scattering will be diminished.

%Therefore, we conclude that upon crystallization from the 2D columnar liquid crystal--whose response is isotropic in the xy plane--frustration induced by the helical symmetry of DNA gives rise to an anisotropic elastic response of a generic form captured by eqs. (\ref{eq.19}) - (\ref{eq.21}).  

What is perhaps most surprising is the relationship between the anisotropic elasticity of the DNA array and the deformations which carry the systems from one frustrated groundstate to another.  It is clear from Fig. \ref{fig.2} that the rhombohedral groundstate is most directly transformed into the higher density orthorhombic groundstate by an {\it elongational shear}, i.e. $u_y= -\epsilon y$ and $u_x= \epsilon x$  [see Fig. \ref{fig.3} (c)].   According the linear theory, such modes lead to particularly large elastic energy costs.  The ratio of the elastic cost of this mode relative to simple shear (i.e. $u_y = \epsilon x$ and $u_x=0$) is $9 (2  - \alpha)/(1-\alpha) \geq 18$, where $\alpha =\Lambda_f^2/(\mu' C_\perp)$.  While the linear-elastic theory of the frustrated rhombohedral DNA crystal does not provide a full description of the structural transition to orthorhombic groundstate, favored at higher densities~\cite{livolant_leforestier}, it nonetheless makes certain generic predictions regarding the dynamics of this phase transition.  Specifically, we expect the rhombohedral DNA crystal must pass through some intermediate state associated with a simple-shear deformation before reaching the orthorhombic state since the crystal is mechanically least stable to these modes.  Interestingly, the orthorhombic state is not accessible by pure simple shear alone, and secondary distortion is required to accomplish this structural transition.  This should be distinguished from scenario where the system bypasses the rhombohedral state and may pass from hexatic to orthorhombic crystal through series of continuous transitions~\cite{lorman}. In light of this prediction, it would interesting to resolve experimentally the structural intermediates which accommodate the observed change in symmetry between these states of DNA packing.

\section{Summary}  
Relying on symmetry considerations alone, we have constructed an elastic theory of DNA crystals, which are inherently frustrated by the underlying helical structure of the constituent molecules.  As the system passes from a hexagonally-ordered liquid crystalline state to a 3D crystal at higher density, the interactions between double-helical molecules are, in general, quite frustrated leading to spontaneously broken discrete symmetry associated with the lower 3-fold symmetry of the groundstate.  A general consequence of the of this broken symmetry is a highly anisotropic response to deformations of in-plane order, with a pronounced resistance to elongational shear modes associated with structural transitions to higher density groundstates.

\acknowledgments
I have benefitted from useful discussions with R. Bruinsma and J. Rudnick on this topic.  This work was supported by the NSF under DMR Grant No. 04-04507 and UMass, Amherst through a Healey Endowment Grant.

\end{document}